\documentclass{article}
\usepackage{amssymb,amsmath,cite,bm,eufrak}
\def\e{\EuFrak{e}}
\def\d{\mathrm{d}}

\def\id{\mathbf{1}}
\def\ir{\mathrm{i}}
\def\tr{\mathrm{tr}}

\begin{document}
\begin{titlepage}
\noindent{\large\textbf{Dirac theory on a space with linear Lie type fuzziness}}

\vspace{\baselineskip}

\begin{center}
{
Ahmad~Shariati~{\footnote{shariati@mailaps.org}}}\\
Mohammad~Khorrami~{\footnote{mamwad@mailaps.org}}\\
Amir~H.~Fatollahi~{\footnote{fath@alzahra.ac.ir}}\\
\vskip 1cm
\textit{ Department of Physics, Alzahra University,
Tehran 1993893973, Iran }
\end{center}

\vspace{\baselineskip}
\begin{abstract}
\noindent A spinor theory on a space with linear Lie type
noncommutativity among spatial coordinates is presented.
The model is based on the Fourier space corresponding to
spatial coordinates, as this Fourier space is commutative.
When the group is compact, the real space exhibits lattice
characteristics (as the eigenvalues of space operators are discrete),
and the similarity of such a \emph{lattice} with ordinary
lattices is manifested, among other things, in a phenomenon
resembling the famous \emph{fermion doubling} problem. A projection
is introduced to make the dynamical number of spinors equal to that
corresponding to the ordinary space. The actions for free and
interacting spinors (with Fermi-like interactions) are
presented. The Feynman rules are extracted and 1-loop corrections
are investigated.
\end{abstract}
\end{titlepage}
\section{Introduction}
In recent years much attention has been paid to the formulation and study
of field theories on noncommutative spaces. The motivation is
partly the natural appearance of noncommutative spaces in some
areas of physics, including recently in string theory. In
particular it has been understood that the longitudinal directions
of D-branes in the presence of a constant $B$-field background
appear to be noncommutative, as seen by the ends of open strings
\cite{9908142,99-2,99-3,99-4}. In this case the coordinates
satisfy the canonical relation
\begin{equation}\label{07.1}
[\hat{x}_\mu,\hat{x}_\nu]=\ir\,\theta_{\mu\,\nu}\,\id,
\end{equation}
in which $\theta$ is an antisymmetric constant tensor and $\id$
is the unit operator. The theoretical and phenomenological
implications of such noncommutative coordinates have been extensively
studied.

One direction to extend studies on noncommutative spaces is to
consider spaces where the commutators of the coordinates are not
constants. Examples of this kind are the noncommutative cylinder
and the $q$-deformed plane (the Manin plane \cite{manin}), the so-called
$\kappa$-Poincar\'{e} algebra \cite{luk} (see also \cite{majid,ruegg,amelino,kappa,chai}),
and linear noncommutativity of the Lie algebra type
\cite{snyder} (see also \cite{wess,sasak}). In the latter the dimensionless spatial
position operators satisfy the commutation relations of a Lie
algebra:
\begin{equation}\label{07.2}
[\hat{x}_a,\hat{x}_b]= f^c{}_{a\, b}\,\hat{x}_c,
\end{equation}
where $f^c{}_{a\,b}$'s are structure constants of a Lie algebra.
One example of this kind is the algebra SO(3), or SU(2). A special
case of this is the so called fuzzy sphere \cite{madore} (see also \cite{presnaj}),
where an irreducible representation of the position operators is
used which makes the Casimir of the algebra,
$(\hat{x}_1)^2+(\hat{x}_2)^2+(\hat{x}_3)^2$, a multiple of the
identity operator (a constant, hence the name sphere). One can
consider the square root of this Casimir as the radius of the
fuzzy sphere. This is, however, a noncommutative version of a
two-dimensional space (sphere).

In \cite{0612013,fakE1,fakE2} a model was introduced in which the
representation was not restricted to an irreducible one, instead
the whole group was employed. In particular the regular
representation of the group was considered, which contains all
representations. As a consequence in such models one is dealing
with the whole space, rather than a sub-space, like the case of
fuzzy sphere as a 2-dimensional surface. In \cite{0612013} basic
ingredients for calculus on a linear fuzzy space, as well as basic
notions for a field theory on such a space, were introduced. In
\cite{fakE1, fakE2} basic elements for calculating the matrix elements
corresponding to transition between initial and final states,
together with the explicit expressions for tree and one-loop
amplitudes were given. It is observed that models based on
Lie algebra type noncommutativity enjoy three features:
\begin{itemize}
\item They are free from any ultraviolet divergences if the group
is compact.
\item There is no momentum conservation in such
theories.
\item In the transition amplitudes only the so-called
planar graphs contribute.
\end{itemize} The reason for latter is that the non-planar graphs
are proportional to $\delta$-distributions whose dimensions are
less than their analogues coming from the planar sector, and so
their contributions vanish in the infinite-volume limit usually
taken in transition amplitudes \cite{fakE2}. One consequence of
a different mass-shell condition of these kinds of theory was explored
in \cite{skf}.

In \cite{kfs} the classical mechanics defined on a space with
SU(2) fuzziness was studied. In particular, the Poisson structure
induced by noncommutativity of SU(2) type was investigated, for
either the Cartesian or Euler parameterization of SU(2) group. The
consequences of SU(2)-symmetry in such spaces on integrability,
were also studied in \cite{kfs}. In \cite{fsk} the quantum mechanics
on a space with SU(2) fuzziness was examined. In particular,
the commutation relations of the position and momentum
operators corresponding to spaces with Lie-algebra
noncommutativity in the configuration space, as well as
the eigen-value problem for the SU(2)-invariant systems were studied.
The consequences of the Lie type noncommutativity of space on thermodynamical
properties have been explored in \cite{shin,fsmjmp}.

The purpose of this work is to develop a spinor theory on a space
with linear Lie type noncommutativity (among spatial coordinates),
specially corresponding to a Lie type noncommutativity of the form
SU(2), in which case the number of spatial coordinates is 3.
The model is basically developed in the Fourier space, as it is
commutative (contrary to the real space). When the group which
corresponds to the noncommutativity is compact, the real space behaves
in some sense like a lattice, as the eigenvalues of the coordinate
operators are discrete. This is manifested, among other things, in
the fact that the dynamical number of fermions is more than the corresponding
number in the real space, similar to the famous \emph{fermion doubling}
problem arisen in fermion theories on ordinary lattices. A projection
is introduced to make the dynamical number of spinors equal to
the number corresponding to the ordinary space.
The actions for free and Fermi-like interacting spinors are
presented. As the momentum space is compact, the interacting
theory is finite and does not suffer from ultraviolet divergences,
contrary to the case of ordinary (commuting) space on which
the theory is ultraviolet divergent, and not renormalizable.
It is seen that for such theories, the 1-loop correction
to the propagator has no non-planar contribution, contrary to
the case of scalar fields on noncommutative spaces \cite{fakE2}.
However, the 1-loop correction to the 4-point function is shown to
get both planar and non-planar contributions.

The scheme of the rest of this paper is the following. In section 2,
a brief introduction of the group algebra is given, mainly to fix
notation. In section 3, the Dirac equation is presented in the
momentum space, first in the case of the commutative space, then
in the case of a noncommutative space. There it is shown that
the number of dynamical degrees of spinors is more than
the corresponding number in the ordinary space, and a projection
is introduced to reduce the number of dynamical degrees of spinors
to that of ordinary space. In section 4 the Dirac action on
a noncommutative space is presented, for free fermions as well as
fermions with Fermi-like interactions. The Feynman rules are extracted,
and 1-loop corrections to the propagator and the 4-point function
are studied. Section 5 is devoted to the concluding remarks.
\section{The group algebra}
Assume that there exists a unique measure $\d U$ (up to a
multiplicative constant) with the invariance properties
\begin{align}\label{07.3}
\d (V\,U)&=\d U,\nonumber\\
\d (U\,V)&=\d U,\nonumber\\
\d (U^{-1})&=\d U,
\end{align}
for any arbitrary element ($V$) of the group.
For a compact group $G$, such a measure does exist. There are,
however, groups which are not compact but for them as well such
a measure exists. Examples are noncompact Abelian groups.

The meaning of (\ref{07.3}), is that the measure is invariant
under the left-translation, right-translation, and inversion.
This measure, the (left-right-invariant) Haar measure, is unique up to a
normalization constant, which defines the volume of the group:
\begin{equation}\label{07.4}
\int_G\d U=\mathrm{vol}(G).
\end{equation}
Using this measure, one constructs a vector space as follows.
Corresponding to each group element $U$ an element $\e(U)$ is
introduced, and the elements of the vector space are linear
combinations of these elements:
\begin{equation}\label{07.5}
f:=\int\d U\;f(U)\,\e(U),
\end{equation}
The group algebra is this vector space, equipped with the
multiplication
\begin{equation}\label{07.6}
f\bullet g:=\int\d U\,\d V\; f(U)\,g(V)\,\e(U\,V),
\end{equation}
where $(U\,V)$ is the usual product of the group elements. $f(U)$
and $g(U)$ belong to a field (here the field of complex numbers).
It can be seen that if one takes the central extension of the
group U(1)$\times\cdots\times$U(1), the so-called Heisenberg
group, with the algebra (\ref{07.1}), the above definition results
in the well-known star product of two functions, provided $f$ and
$g$ are interpreted as the Fourier transforms of the functions.

So there is a correspondence between functionals defined on the
group, and the group algebra. The definition (\ref{07.6}) can be
rewritten as
\begin{equation}\label{07.7}
(f\bullet g)(W)=\int\d V\;f(W\,V^{-1})\,g(V).
\end{equation}

The delta distribution is defined through
\begin{equation}\label{07.8}
\int\d U\;\delta(U)\,f(U):=f(\id),
\end{equation}
where $\id$ is the identity element of the group.

Next, one can define an inner product on the group algebra.
Defining
\begin{equation}\label{07.9}
\langle\e(U),\e(V)\rangle:=\delta(U^{-1}\,V),
\end{equation}
and demanding that the inner product be linear with respect to its
second argument and antilinear with respect to its first argument,
one arrives at
\begin{equation}\label{07.10}
\langle f,g\rangle=\int\d U\;f^*(U)\,g(U).
\end{equation}

Finally, one defines a star operation through
\begin{equation}\label{07.11}
f^\star(U):=f^*(U^{-1}).
\end{equation}
This is in fact equivalent to definition of the star operation in
the group algebra as
\begin{equation}\label{07.12}
[\e(U)]^\star:=\e(U^{-1}).
\end{equation}
It is then easy to see that
\begin{align}\label{07.13}
(f\,g)^\star=&g^\star\,f^\star,\\ \label{07.14} \langle f, g\rangle=&
(f^\star\,g)(\id).
\end{align}
\section{The Dirac equation}
To write the Dirac equation on a noncommutative space, let us begin
with the Dirac equation on a commutative space in terms of the Fourier
transform. For the 4 dimensional space-time, the following conventions
are used,
\begin{align}\label{07.15}
\{\gamma^\sigma,\gamma^\rho\}&=2\,\eta^{\sigma\,\rho},\\ \label{07.16}
\gamma^0&=\ir\,\beta,\\ \label{07.17}
\bar\psi&=\psi^\dagger\,\beta,
\end{align}
where $\eta$ is the Minkowski metric with the signature
$(-,+,+,+)$.
\subsection{The Dirac equation in the Fourier space}
The Dirac equation in the Fourier space on a commutative space is
\begin{equation}\label{07.18}
(\gamma^0\,\partial_0+\ir\,\gamma^a\,k_a-\mu)\,\psi(t,U)=0,
\end{equation}
where
\begin{align}\label{07.19}
U&:=\exp(\ell\,k^a\,T_a),\\ \label{07.20}
k_a&:=\delta_{a\,b}\,k^b,
\end{align}
and
\begin{align}\label{07.21}
\tr(T_a\,T_b)&=c\,\delta_{a\,b},\\ \label{07.22}
\tr(T_a)&=0.
\end{align}
$\ell$ is a parameter of dimension length, and $T_a$'s are
generators of some group, in some representation.
But as long as (\ref{07.18}) is studied, it is not
important what the value of $\ell$ is, and what the group is,
provided the dependence of the group element $U$ on $\bm{k}$
is one to one. If the latter condition is violated, then
sending $\ell$ to zero effectively makes the dependence of
$U$ on $\bm{k}$ one to one.
It is then seen that the Dirac equation can be written as
\begin{equation}\label{07.23}
\{\gamma^0\,\partial_0+\ir\,c^{-1}\,\gamma^a\,\lim_{\ell\to 0}[\ell^{-1}\,\tr(T_a\,U)]
-\mu\}\,\psi(t,U)=0.
\end{equation}
\subsection{The Dirac equation on a noncommutative space}
Equation (\ref{07.23}) provides one with a way of writing the Dirac equation
on a noncommutative space. This is done essentially by removing the limit
$\ell\to 0$, and taking (as usual) $T_a$'s to be the generator of a
Lie group $G$. So the equation reads
\begin{equation}\label{07.24}
[\mathcal{D}(U)]\,\psi(t,U)=0,
\end{equation}
where the Dirac operator $\mathcal{D}$ is defined as
\begin{equation}\label{07.25}
\mathcal{D}(U):=\gamma^0\,\partial_0+\ir\,c^{-1}\,\gamma^a\,\ell^{-1}\,\tr(T_a\,U)-\mu.
\end{equation}
Using
\begin{equation}\label{07.26}
\tr(T_a\,U)=\ell^{-1}\,\frac{\partial\tr(U)}{\partial k^a},
\end{equation}
one obtains
\begin{equation}\label{07.27}
\mathcal{D}(U)=\gamma^0\,\partial_0+\ir\,c^{-1}\,\gamma^a\,\ell^{-2}\,
\frac{\partial\tr(U)}{\partial k^a}-\mu.
\end{equation}
Of course $c$ and $\tr(U)$ depend on the representation.
\subsection{The Dirac equation for the gauge group SU(2)}
For the group SU(2), one has for the spin $s$ representation
\begin{equation}\label{07.28}
\tr(U)=\frac{\displaystyle{\sin\left[\ell\,k\,\left(s+\frac{1}{2}\right)\right]}}
{\displaystyle{\sin\frac{\ell\,k}{2}}},
\end{equation}
where
\begin{equation}\label{07.29}
k:=\sqrt{\delta_{a\,b}\,k^a\,k^b}.
\end{equation}
So,
\begin{equation}\label{07.30}
\ell^{-1}\,\frac{\partial\tr(U)}{\partial k^a}=\frac{k_a}{k}\,
\frac{s\,\sin[(s+1)\,\ell\,k]-(s+1)\,\sin(s\,\ell\,k)}
{2\,\sin^2(\ell\,k/2)}.
\end{equation}
One also has
\begin{equation}\label{07.31}
c=-\frac{s\,(s+1)\,(2\,s+1)}{3}.
\end{equation}
So one obtains for the Dirac operator
\begin{equation}\label{07.32}
\mathcal{D}(U)=\gamma^0\,\partial_0-\frac{3\,\{s\,\sin[(s+1)\,\ell\,k]-(s+1)\,\sin(s\,\ell\,k)\}}
{s\,(s+1)\,(2\,s+1)[2\,\ell\,k\,\sin^2(\ell\,k/2)]}\,\,(\ir\,\gamma^a\,k_a)-\mu.
\end{equation}

In the special case $s=1/2$, this becomes
\begin{equation}\label{07.33}
\mathcal{D}(U)=\gamma^0\,\partial_0+\frac{2\,\sin(\ell\,k/2)}{\ell\,k}\,(\ir\,\gamma^a\,k_a)-\mu.
\end{equation}
The mass shell condition can be obtained, similar to the case of
commutative spaces, by multiplying the Dirac equation from the left
by the conjugate operator. The result would be
\begin{align}\label{07.34}
0&=\left[\gamma^0\,\partial_0+\frac{2\,\sin(\ell\,k/2)}{\ell\,k}\,(\ir\,\gamma^a\,k_a)+\mu\right]
\nonumber\\&\quad\times
\left[\gamma^0\,\partial_0+\frac{2\,\sin(\ell\,k/2)}{\ell\,k}\,(\ir\,\gamma^a\,k_a)-\mu\right]
\,\psi(t,U),\nonumber\\&=\left\{-(\partial_0)^2-
\left[\frac{2\,\sin(\ell\,k/2)}{\ell}\right]^2-\mu^2\right\}\,\psi(t,U),
\end{align}
which results in the following mass shell condition
\begin{equation}\label{07.35}
\omega^2=\frac{2\,[1-\cos(\ell\,k)]}{\ell^2}+\mu^2.
\end{equation}
For the group SU(2), the range of $k$ to cover all of the group
once, is
\begin{equation}\label{07.36}
0\leq(\ell\,k)\leq(2\,\pi).
\end{equation}
So the energy is clearly not an increasing function
of $k$. It is so for $(\ell\,k)$ between $0$ and $\pi$.
It seems that the whole range of $k$ produces two
copies of the spinor field. That is similar to what
arises in the context of spinor fields on regular lattices,
the so-called \emph{fermion doubling} problem \cite{rothe}.
There, corresponding to each direction there are two
copies of the spinor field. So that corresponding to
a four dimensional lattice there are 16 copies of
the spinor field. One could get rid of the additional
(redundant) fields, by introducing suitable projections
which commute with the equation of motion operator, so
that different eigenvectors of the projections satisfy
the equation separately. This approach is similar to
the momentum space formulation of the so-called
\emph{staggered fermions} in ordinary lattice gauge theories
\cite{rothe}. For the present case, one
notices that changing $(\ell\,k_a)$ to $[(\ell\,k_a)-(2\,\pi\,k_a/k)]$
is equivalent to changing $U$ to $(-U)$. Such a transformation
changes $(\ell\,k)$ to $|\ell\,k-2\,\pi|$, so it turns the region 
$(\ell\,k)\in[0,\pi]$ into $(\ell\,k)\in[\pi,2\,\pi]$,
and vice versa. So a possible projection could
be constructed through the operator $\mathcal{M}$ with
\begin{equation}\label{07.37}
(\mathcal{M}\,\psi)(t,U):=M\,\psi(t,-U),
\end{equation}
which results in
\begin{equation}\label{07.38}
(\mathcal{M}\,\mathcal{D}\,\mathcal{M}^{-1})(U):=M\,[\mathcal{D}(-U)]\,M^{-1}.
\end{equation}
Multiplying $\mathcal{M}$ from left on (\ref{07.24}), one arrives at
\begin{equation}\label{07.39}
\{M\,[\mathcal{D}(-U)]\,M^{-1}\}\,(\mathcal{M}\,\psi)(t,U)=0.
\end{equation}
So $(\mathcal{M}\,\psi)$ satisfies the same equation $\psi$ satisfies,
provided
\begin{equation}\label{07.40}
\{M\,[\mathcal{D}(-U)]\,M^{-1}\}=\mathcal{D}(U),
\end{equation}
or
\begin{align}\label{07.41}
M\,\gamma^0\,M^{-1}&=\gamma^0,\nonumber\\
M\,\gamma^a\,M^{-1}&=-\gamma^a,
\end{align}
which shows that $M$ is proportional to $\beta$.
Taking it to be the same as $\beta$, it is seen that
\begin{equation}\label{07.42}
\mathcal{M}^2=1.
\end{equation}
So the eigenvalues of $\mathcal{M}$ are $\pm 1$.
One can then decompose the spinor field as
\begin{equation}\label{07.43}
\psi=\psi^++\psi^-,
\end{equation}
where
\begin{equation}\label{07.44}
\psi^\pm:=\frac{1\pm\mathcal{M}}{2}\,\psi.
\end{equation}
Obviously the equations for $\psi^+$ and $\psi^-$ are decoupled.
So that one can take only one of the spinor fields, say $\psi^-$,
and write the equation for that. This way there remains only
one copy of the spinor field, and one effectively needs only
the values of $k$ corresponding to $(\ell\,k)$ not
larger than $\pi$, as $\psi^-(t,-U)$ is not
independent of $\psi^-(t,U)$.

It is obvious that the above construction works for any group
with the property that if $U$ belongs to the group, $(-U)$
belongs to the group as well. Let's call such a group a
double copy group. One also notes that $\mathcal{M}$ is not
the parity operator. The parity operator transforms $k$ to
$(-k)$, which is equivalent to transforming $U$ to
$U^{-1}$, while $\mathcal{M}$ transforms $U$ to $(-U)$.

\section{The Dirac action on a noncommutative space}
The Dirac equation (\ref{07.24}) for a free fermion
can be obtained from an action.
\subsection{The Dirac action for a free field}
The Dirac action for a free field on a noncommutative space is written as
\begin{equation}\label{07.45}
S_\mathrm{free}=\frac{1}{\sigma}\,\int\d t\int\d U\;\bar\psi(t,U^{-1})\,[\mathcal{D}(U)]\,\psi(t,U),
\end{equation}
where $\sigma$ is a symmetry factor:
\begin{equation}\label{07.46}
\sigma=\begin{cases}
1,&\mbox{$G$ is not a double copy group}\\
2,&\mbox{$G$ is a double copy group}
\end{cases},
\end{equation}
and the Haar measure $\d U$ is normalized so that
\begin{equation}\label{07.47}
\d U\sim\frac{\d^D k}{(2\,\pi)^D},\qquad U\sim\id.
\end{equation}
The symmetry factor ensures that in the limit $(\ell\to 0)$,
the commutative action is recovered with proper normalization.

Obviously, if $G$ is a double copy group one has
\begin{equation}\label{07.48}
\overline{(\mathcal{M}\,\psi)}(t,U)=[\bar\psi(t,-U)]\,M^{-1},
\end{equation}
which together with
\begin{equation}\label{07.49}
\d(-U)=\d U,
\end{equation}
shows that the action $S_\mathrm{free}$ enjoys the following symmetry
\begin{equation}\label{07.50}
S_\mathrm{free}(\mathcal{M}\,\psi)=S_\mathrm{free}(\psi).
\end{equation}

Using the action (\ref{07.45}), the propagator
(in the full Fourier space) is found to be
\begin{equation}\label{07.51}
\check\Delta(\omega,U)=(\ir\,\hbar)\,[\check{\mathcal{D}}(\omega,U)]^{-1},
\end{equation}
where
\begin{equation}\label{07.52}
\check{\mathcal{D}}(\omega,U)=-\ir\,\gamma^0\,\omega+\ir\,c^{-1}\,\gamma^a\,\ell^{-1}\,\tr(T_a\,U)-\mu.
\end{equation}
so,
\begin{align}\label{07.53}
\check\Delta(\omega,U)&=(\ir\,\hbar)\,
\left[-\ir\,\gamma^0\,\omega+\ir\,c^{-1}\,\gamma^a\,\ell^{-1}\,\tr(T_a\,U)-\mu\right]^{-1},\nonumber\\
&=\frac{\ir\,\hbar}{\omega^2-(c\,\ell)^{-2}\,\tr(T_b\,U)\,\tr(T^b\,U)-\mu^2}\nonumber\\
&\quad\times\left[-\ir\,\gamma^0\,\omega+\ir\,c^{-1}\,\gamma^a\,\ell^{-1}\,\tr(T_a\,U)+\mu\right].
\end{align}
For the group SU(2), the action (\ref{07.45}) is reduced to
\begin{align}\label{07.54}
S_\mathrm{free}&=\frac{1}{2}\,\int\d t\int\d U\;\bar\psi(t,U^{-1})\,
[\mathcal{D}(U)]\,\psi(t,U),\nonumber\\
&=\frac{1}{2}\,\int\d t\int\d U\;\bar\psi^-(t,U^{-1})\,
[\mathcal{D}(U)]\,\psi^-(t,U)\nonumber\\
&\quad+\frac{1}{2}\,\int\d t\int\d U\;\bar\psi^+(t,U^{-1})\,
[\mathcal{D}(U)]\,\psi^+(t,U),\nonumber\\
&=:S^-_\mathrm{free}+S^+_\mathrm{free},
\end{align}
where $\mathcal{D}(U)$ is of the form (\ref{07.33}).
The propagator would be
\begin{align}\label{07.55}
\check\Delta(\omega,U)&=(\ir\,\hbar)\,
\left[-\ir\,\gamma^0\,\omega+\frac{2\,\sin(\ell\,k/2)}{\ell\,k}\,(\ir\,\gamma^a\,k_a)-\mu\right]^{-1},
\nonumber\\
&=\frac{\ir\,\hbar}{\omega^2-(4/\ell^2)\,\sin^2(\ell\,k/2)-\mu^2}\,
\left[-\ir\,\gamma^0\,\omega+\frac{2\,\sin(\ell\,k/2)}{\ell\,k}\,(\ir\,\gamma^a\,k_a)+\mu\right].
\end{align}
The action (\ref{07.54}) contains two copies of the fermion field, as
previously explained. So the proper action of a single free fermion field would
be $S^-_\mathrm{free}$ (or $S^+_\mathrm{free}$).
\subsection{Interacting Dirac fields}
An example for the interaction of Dirac fields is a Fermi like
interaction, corresponding to the action
\begin{align}\label{07.56}
S_\mathrm{Fermi}&=-\frac{g}{(j!)^2}\,\int\d t \int\d U_1\cdots\d U_{2\,j}\;
\delta(U_1\cdots U_{2\,j})\nonumber\\
&\qquad\times[\bar\psi(t,U_1)\,\psi(t,U_2)]\cdots[\bar\psi(t,U_{2\,j-1})\,\psi(t,U_{2\,j})].
\end{align}
This interaction is not renormalizable in the ordinary space. But here
there are no ultraviolet divergences (as far as the group is compact).
Again, if $G$ is a double copy group one could write actions which
contain only one copy of the fermion field:
\begin{align}\label{07.57}
S^-_\mathrm{Fermi}&=-\frac{g}{\sigma^{2\,j-1}\,(j!)^2}\,\int\d t \int\d U_1\cdots\d U_{2\,j}\;
\delta(U_1\cdots U_{2\,j})\nonumber\\
&\qquad\times[\bar\psi^-(t,U_1)\,\psi^-(t,U_2)]\cdots[\bar\psi^-(t,U_{2\,j-1})\,\psi^-(t,U_{2\,j})].
\end{align}
The full action, would then be
\begin{align}\label{07.58}
S^-&=\frac{1}{\sigma}\,\int\d t\int\d U_1\,\d U_2\;\delta(U_1\,U_2)\,\bar\psi^-(t,U_1)\,
[\mathcal{D}(U_2)]\,\psi^-(t,U_2)\nonumber\\
&\quad-\frac{g}{\sigma^{2\,j-1}\,(j!)^2}\,\int\d t \int\d U_1\cdots\d U_{2\,j}\;
\delta(U_1\cdots U_{2\,j})\nonumber\\
&\qquad\times[\bar\psi^-(t,U_1)\,\psi^-(t,U_2)]\cdots[\bar\psi^-(t,U_{2\,j-1})\,\psi^-(t,U_{2\,j})].
\end{align}
The vertex corresponding to such an interaction reads
\begin{align}\label{07.59}
\mathcal{V}^{\alpha_1\,\alpha_3\cdots\alpha_{2\,j-1}}_{\alpha_2\,\alpha_4\cdots\alpha_{2\,j}}
(U_1,\dots,U_{2\,j})&=\frac{1}{\sigma^{2\,j-1}\,(j!)^2}\,\frac{g}{\ir\,\hbar}\,2\pi\,
\delta(\omega_1+\cdots+\omega_{2\,j})\nonumber\\
&\quad\times
\sum_{\Pi,\Pi'}\zeta_\Pi\,\zeta_{\Pi'}\,\delta^{\alpha_{2\,\Pi(1)-1}}_{\alpha_{2\,\Pi'(1)}}
\cdots\delta^{\alpha_{2\,\Pi(j)-1}}_{\alpha_{2\,\Pi'(j)}}\nonumber\\
&\quad\times\delta(U_{2\,\Pi(1)-1}\,U_{2\,\Pi'(1)}\cdots U_{2\,\Pi(j)-1}\,U_{2\,\Pi'(j)}),
\end{align}
where $\Pi$ and $\Pi'$ are $j$-permutations, and $\zeta_\Pi$ is
the sign of the permutation $\Pi$ (plus one for even permutations,
and minus one for odd permutations).

For the 4-fermion interaction, the above would be
\begin{align}\label{07.60}
\mathcal{V}^{\alpha_1\,\alpha_3}_{\alpha_2\,\alpha_4}
(U_1,\dots,U_{4})&=\frac{1}{2\sigma^{3}}\,\frac{g}{\ir\,\hbar}\,2\pi\,
\delta(\omega_1+\cdots+\omega_{4})\nonumber\\
&\quad\times[
\delta^{\alpha_1}_{\alpha_2}\,\delta^{\alpha_3}_{\alpha_4}\,\delta(U_1\,U_2\,U_3\,U_4)
-\delta^{\alpha_1}_{\alpha_4}\,\delta^{\alpha_3}_{\alpha_2}\,\delta(U_1\,U_4\,U_3\,U_2)],
\end{align}
where use has been made of the fact that
\begin{equation}\label{07.61}
\delta(U\,U')=\delta(U'\,U).
\end{equation}
and so forth. It is seen that the above vertex respects a
deformed momentum conservation. In commutative space,
the momentum conservation is that the sum of all momenta
should vanish. In noncommutative space, however, processes
are allowed for which the product of group elements are unit,
and as different orderings in the products are possible,
there could be several conservation delta functions in the vertex,
which are not the same. In the above, for example, there are two
different delta functions. A similar thing occurs in theories
defined on $\kappa$-deformed spaces. In these theories,
the ordinary summation of momenta in each
vertex is replaced by a new rule of summation, occasionally
called as doted-sum ($\dot{+}$) \cite{amelino}. This new sum, in
contrast to the ordinary sum, is non-Abelian, and as a
consequence, the delta functions's corresponding to different
possible orderings of legs are different \cite{amelino,kappa}.
\subsection{1-loop correction of the 2-point function}
The 2-point function has two external legs 1 and 2. The 1-loop
correction is simply the vertex-function (\ref{07.60}), contracted
with the propagator corresponding to the legs 3 and 4:
\begin{align}\label{07.62}
\Gamma^{(2)}_1{}^{\alpha_1}_{\alpha_2}&=\int\d U_3\,\d U_4\,\frac{\d\omega_3}{2\,\pi}
\,\frac{\d\omega_4}{2\,\pi}\;\delta(U_3\,U_4)\,(2\,\pi)\,\delta(\omega_3+\omega_4)
\,\mathcal{V}^{\alpha_1\,\alpha_3}_{\alpha_2\,\alpha_4}\,
\check\Delta^{\alpha_4}{}_{\alpha_3}(\omega_3,U_3)
\nonumber\\
&=\frac{g}{\ir\,\hbar}\frac{1}{2\,\sigma^{3}}\,(2\,\pi)\,\delta(\omega_1+\omega_2)\,
\delta(U_1\,U_2)\,\int\d U_3\,\frac{\d\omega_3}{2\pi}\;\frac{\ir\,\hbar}{\omega_3^2 +
O(U_3)}\nonumber\\
&\quad\times\{\delta^{\alpha_1}_{\alpha_2}\,\tr[\check{\mathcal{D}}'(\omega_3,U_3)]- \check{\mathcal{D}}'^{\alpha_1}{}_{\alpha_2}(\omega_3,U_3)\},
\end{align}
where
\begin{align}\label{07.63}
\check{\mathcal{D}}'(\omega,U)&:=-\ir\,\gamma^0\,\omega+\ir\,c^{-1}\,\gamma^a\,\ell^{-1}\,\tr(T_a\,U)+\mu,
\nonumber\\
O(U)&:=-(c\,\ell)^{-2}\,\tr(T_b\,U)\,\tr(T^b\,U)-\mu^2.
\end{align}
The coefficient $\delta(U_1\,U_2)$ shows that for the propagator,
up to 1-loop correction the analog of momentum conservation
still holds.

The above expression for 2-point function may be contrasted with
the similar one for the scalar fields \cite{fakE2}. In the case for scalars,
there is a term in which the $\delta$-function consists the loop variable,
and so could not be brought out the integral. In that case, such a term
is called as the non-planar contribution. Here, for spinors, we see that
such a term is absent, and the contribution is totally planar.
\subsection{1-loop correction of the 4-point function}
The aim of this subsection is to investigate the possibility
of non-planar contributions. Labeling the external legs 1 through 4,
it is seen that there are 6 distinct combinations of
pseudo-conservation terms:
\begin{align}\label{07.64}
\mathcal{C}_\mathrm{I}&=\delta(U_1\,U_2\,U_5\,U_6)\,\delta(U_3\,U_4\,U_6^{-1}\,U_5^{-1}),\nonumber\\
\mathcal{C}_\mathrm{II}&=\delta(U_1\,U_2\,U_5\,U_6)\,\delta(U_3\,U_5^{-1}\,U_6^{-1}\,U_4),\nonumber\\
\mathcal{C}_\mathrm{III}&=\delta(U_1\,U_4\,U_5\,U_6)\,\delta(U_3\,U_2\,U_6^{-1}\,U_5^{-1}),\nonumber\\
\mathcal{C}_\mathrm{IV}&=\delta(U_1\,U_4\,U_5\,U_6)\,\delta(U_3\,U_5^{-1}\,U_6^{-1}\,U_2),\nonumber\\
\mathcal{C}_\mathrm{V}&=\delta(U_1\,U_6\,U_3\,U_5^{-1})\,\delta(U_5\,U_2\,U_6^{-1}\,U_4),\nonumber\\
\mathcal{C}_\mathrm{VI}&=\delta(U_1\,U_6\,U_3\,U_5^{-1})\,\delta(U_5\,U_4\,U_6^{-1}\,U_2),
\end{align}
where the labels 5 and 6 refer to internal legs. Integrating over $U_6$,
the corresponding contributions become
\begin{align}\label{07.65}
\mathcal{C}'_\mathrm{I}&=\delta(U_1\,U_2\,U_3\,U_4),\nonumber\\
\mathcal{C}'_\mathrm{II}&=\delta(U_1\,U_2\,U_5\,U_4\,U_3\,U_5^{-1}),\nonumber\\
\mathcal{C}'_\mathrm{III}&=\delta(U_1\,U_4\,U_3\,U_2),\nonumber\\
\mathcal{C}'_\mathrm{IV}&=\delta(U_1\,U_4\,U_5\,U_2\,U_3\,U_5^{-1}),\nonumber\\
\mathcal{C}'_\mathrm{V}&=\delta(U_1\,U_4\,U_5\,U_2\,U_3\,U_5^{-1}),\nonumber\\
\mathcal{C}'_\mathrm{VI}&=\delta(U_1\,U_2\,U_5\,U_4\,U_3\,U_5^{-1}).
\end{align}
It is seen that of these six channels, the first and the third
correspond to planar contributions, while others correspond to
non-planar ones.

\section{Concluding remarks}
A spinor theory on a space with linear Lie type noncommutativity
among spatial coordinates was presented. It was shown that
the dynamical number of spinors could be more than the number
corresponding to the commutative spaces, as a result of
the lattice-like nature of the noncommutative space. This is
similar to the famous \emph{fermion doubling} problem arisen in
the case of regular lattices. A projection was introduced to
remove the additional degrees of freedom. Actions for free and
Fermi-like interacting spinors were presented, and were specialized
to the case where the group corresponding to the noncommutativity
is SU(2). The Feynman rules were extracted and 1-loop corrections
to the 2- and 4-point functions were studied. It was shown that
up to 1-loop, there is no non-planar contribution in the 2-point function,
while there are planar as well as non-planar contributions in the 4-point function.
\\[\baselineskip]
\textbf{Acknowledgement}:  This work was
supported by the Research Council of the Alzahra University.

\end{document}